\documentclass[showpacs,aps,graphicx,twocolumn]{revtex4}
\usepackage{graphicx}

\begin{document}

\title{Efficient entanglement  concentration for three-photon W states with parity check measurement}

\author{Yu-Bo Sheng,$^{1,2,4}$\footnote{Email address:
shengyb@njupt.edu.cn} Lan Zhou,$^3$  Sheng-Mei Zhao,$^{1,4}$ }
\address{$^1$ Institute of Signal Processing  Transmission, Nanjing
University of Posts and Telecommunications, Nanjing, 210003,  China\\
$^2$College of Telecommunications \& Information Engineering,
Nanjing University of Posts and Telecommunications,  Nanjing,
210003,
China\\
 $^3$Beijing National Laboratory for Condensed Matter Physics, Institute of Physics,\\
Chinese Academy of Sciences, Beijing 100190, China\\
 $^4$Key Lab of Broadband Wireless Communication and Sensor Network
 Technology,
 Nanjing University of Posts and Telecommunications, Ministry of
 Education, Nanjing, 210003,
 China\\}

\begin{abstract}
We present an optimal entanglement concentration ECP for an
arbitrary less-entangled W state.  By two of the parties say Alice
and Charlie performing one parity check measurements, they can
obtain the maximally entangled W state with a certain probability.
Otherwise, they can obtain another lesser-entangled W state with
another probability, which can be used to reconcentrated into a
maximally entangled W state. By iterating this ECP  several times,
it has the maximal success probability. This ECP maybe an optimal
one and is useful in current quantum information processing.
\end{abstract}
 \maketitle

\section{introduction}

Entanglement is an essential role in quantum information processing
\cite{computation1,computation2,rmp}.  Quantum teleportation
\cite{teleportation,cteleportation}, quantum dense coding
\cite{densecoding}, quantum key distribution protocols(QKD)
\cite{Ekert91}, quantum secret sharing \cite{QSS1,QSS2,QSS3},
quantum secure direct communication \cite{QSDC1,QSDC2,QSDC3} and
other quantum information processing \cite{QSTS1,QSTS2,QSTS3}
 all need to set up the entanglement channel. However, in practice,
a quantum channel which is to  distribute the photon pairs, is
usually noisy. The noisy channel will degrade the entanglement, and
make the maximally entangled state become a mixed entangled state or
a less-entangled state. Thus, it is of vice important that the
communication parties say Alice and Bob share the maximally
entangled states through such noisy channels. Entanglement
purification and concentration are two powerful tools which were
proposed for that purpose.

Entanglement purification was widely studied since Bennett \emph{et
al.} proposed the entanglement protocol in 1996 based on the
controlled-not (CNOT) gate \cite{C.H.Bennett1,D. Deutsch,M. Murao,M.
Horodecki,Pan1,Simon,Pan2,Yong,sangouard1,wangc1,sangouard2,lixhepp,wangc2,wangc3,shengpra,dengonestep}.
 By local
operations and classical communications (LOCC), it can extract some
high-fidelity entangled systems from a less-entangled ensemble in a
mixed state \cite{C.H.Bennett1}. Entanglement concentration, which
will be explained later, is to distill some maximally entangled
states from an ensemble in a less-entangled pure state
\cite{C.H.Bennett2,swapping1,swapping2,zhao1,zhao2,Yamamoto1,Yamamoto2,bose,cao1,cao2,shengpra2,shengpra3,shengqic,wangxb,shengpla,zhang,dengconcentration}.
In 1996, Bennett \emph{et al.} proposed an original entanglement
concentration protocol (ECP), called the Schmidit projection method
\cite{C.H.Bennett2}. An ECP based on the quantum swapping was
proposed and it was developed by Shi \emph{et al.}
\cite{swapping1,swapping2}. In 2001, Yamamoto \emph{et al.} and Zhao
\emph{et al.} proposed two ECPs based on the linear optics
independently \cite{zhao1,zhao2,Yamamoto1,Yamamoto2}. Both ECPs were
realized experimentally \cite{zhao2,Yamamoto2}. ECPs based on the
cross-Kerr nonlinearity were proposed \cite{shengpra2,shengqic}. In
2011, Wang \emph{et al.} proposed an efficient ECP based on the
quantum dot in an optical cavity \cite{wangc1}.  In 2012, we
proposed an efficient ECP with the help of single photons
\cite{shengpra3}.
 The ECPs described above in optical system are all
focused on the two-particle Bell-state. They are all used to convert
a less-entangled state
$\alpha|H\rangle|H\rangle+\beta|V\rangle|V\rangle$ into a maximally
entangled state
$\frac{1}{\sqrt{2}}(|H\rangle|H\rangle+|V\rangle|V\rangle)$. Here
$|H\rangle$ and $|V\rangle$ represent the horizontal and the
vertical polarizations of photons. Unusually, this kind of ECPs can
be extended to concentrate the multipartite less-entangled GHZ state
$\alpha|H\rangle|H\rangle|H\rangle+\beta|V\rangle|V\rangle|V\rangle$
into a maximally entangled GHZ state
$\frac{1}{\sqrt{2}}(|H\rangle|H\rangle|H\rangle+|V\rangle|V\rangle|V\rangle)$.
Unfortunately, they cannot concentrate the less-entangled W state
$\alpha|V\rangle|H\rangle|H\rangle+\beta|H\rangle|V\rangle|H\rangle+\gamma|H\rangle|H\rangle|V\rangle$
into a maximally entangled W state
$\frac{1}{\sqrt{3}}(|V\rangle|H\rangle|H\rangle+|H\rangle|V\rangle|H\rangle+|H\rangle|H\rangle|V\rangle)$
for the W state cannot be converted into a GHZ state in LOCC
\cite{dur}. In 2003, Cao and Yang proposed an ECP for W class state
with the help of joint unitary transformation \cite{cao}. In 2007,
Zhang \emph{et al.} proposed an ECP based on the Bell- state
measurement \cite{zhanglihua}. In 2010, an ECP for a special W state
$\alpha|H\rangle|H\rangle|V\rangle  +
\beta(|H\rangle|V\rangle|H\rangle + |V\rangle| H\rangle|H\rangle)$
was proposed \cite{wanghf1}. They also proposed an ECP for
three-atom W state in cavity QED system with the same idea
\cite{wanghf2}. Yildiz proposed an optimal distillation of
three-qubit asymmetric W states. In 2011, we also proposed an ECP
for W state resorting to the cross-Kerr nonlinearity and some
special polarized single photons.

In this paper, we describe an efficient ECP which is focused on an
arbitrary three-photon W state. We concentrate an arbitrary
less-entangled W state
$\alpha|V\rangle|H\rangle|H\rangle+\beta|H\rangle|V\rangle|H\rangle+\gamma|H\rangle|H\rangle|V\rangle$
into a maximally entangled W state
$\frac{1}{\sqrt{3}}(|V\rangle|H\rangle|H\rangle+|H\rangle|V\rangle|H\rangle+|H\rangle|H\rangle|V\rangle)$,
with the help of two same conventional single photons of the form
$\frac{1}{\sqrt{2}}(|H\rangle+|V\rangle)$. This ECP resorts the
cross-Kerr nonlinearity to perform the parity check measurement. By
two of the parities performing this ECP one time, they can obtain a
maximally entangled W state with a certain success probability.
Compared with the other ECPs for W state, it has several advantages.
First, it can concentrate an arbitrary less-entangled W state and
conventional ECPs unusually focuse on some special W states. Second,
it only requires the conventional single photons, and other ECPs all
need two copies of less-entangled pairs or some special polarized
single photons. Third, the cross-Kerr nonlinearity can work on the
weak area, and we do not require the large phase shift.  This
advantage makes this protocol more feasible in current technology.
Fourth, this ECP can be repeated to get a higher success
probability.

This paper is organized as follows: in Sec. II, we first describe
the basic element for this ECP, say, the parity check measurement
(PCM) constructed by cross-Kerr nonlinearity. In Sec. III, we
explain this ECP. In Sec. IV, we calculate the entanglement
transformation efficiency for our ECP. Finally, in Sec. V, we
present a discussion and summary.

\section{parity check measurement}
 Before we start this ECP, we first introduce the key element
PCM constructed by cross-Kerr nonlinearity. The cross-Kerr
nonlinearity has been widely studied in current quantum information
processing, such as construction of CNOT gate \cite{QND1}, making a
Bell-state measurement \cite{QND2,shengbellstateanalysis},
performing the entanglement purification and conentration
\cite{shengpra,shengpra2,shengpra3,shengqic}, and so on
\cite{qubit1,qubit2,qubit3,he1,he2,he3,zhangshou,lin1,lin2}. From
Fig. 1, the Hamiltonian of a cross-Kerr nonlinearity is
\cite{QND1,QND2}
\begin{eqnarray}
H=\hbar \chi a^{\dagger}_{s}a_{s}a^{\dagger}_{p}a_{p}.
\end{eqnarray}
Here $\chi$ is the coupling strength of the nonlinearity.
$a^{\dagger}_{s}$ and $a^{\dagger}_{p}$ are the creation operations,
and $a_{s}$ and $a_{p}$ are the destruction operations. We consider
two single photon states
$|\psi_{1}\rangle=c_{1}|H\rangle+d_{1}|V\rangle$ and
$|\psi_{2}\rangle=c_{2}|H\rangle+d_{2}|V\rangle$ coupled with a
coherent state $|\alpha\rangle$ shown in Fig. 1. Here
$|c_{1}|^{2}+|d_{1}|^{2}=1$, and $|c_{2}|^{2}+|d_{2}|^{2}=1$. The
state $|\psi_{1}\rangle$ is in the spatial mode $a_{1}$ and
$|\psi_{2}\rangle$ is in the spatial mode $a_{2}$. The whole system
can be written as:
\begin{eqnarray}
&&|\psi_{1}\rangle\otimes|\psi_{2}\rangle\otimes|\alpha\rangle
=(c_{1}|H\rangle+d_{1}|V\rangle)\nonumber\\
&\otimes&(c_{2}|H\rangle+d_{2}|V\rangle)|\alpha\rangle\nonumber\\
&=&(c_{1}c_{2}|H\rangle|H\rangle+c_{1}d_{2}|H\rangle|V\rangle\nonumber\\
&+&d_{1}c_{2}|V\rangle|H\rangle+d_{1}d_{2}|V\rangle|V\rangle)|\alpha\rangle\nonumber\\
&\rightarrow&(c_{1}c_{2}|H\rangle|H\rangle+d_{1}d_{2}|V\rangle|V\rangle)|\alpha\rangle\nonumber\\
&+&(c_{1}d_{2}|H\rangle|V\rangle |\alpha
e^{-2i\theta}\rangle+d_{1}c_{2}|V\rangle|H\rangle|\alpha
e^{2i\theta}\rangle).\label{couple}
\end{eqnarray}
From Eq. (\ref{couple}), it is obvious that the even parity states
$|H\rangle|H\rangle$ and $|V\rangle|V\rangle$ lead to the coherent
state pick up no phase shift. The odd parity state
$|H\rangle|V\rangle$ leads to the coherent state pick up $-2\theta$
phase shift and $|V\rangle|H\rangle$ picks up the $2\theta$ phase
shift. With an $X$ quadrature measurement in which the states
$|\alpha e^{\pm2\theta}\rangle$ cannot be distinguished, one can
distinguish the even parity state from the odd parity state
according to the different phase shift in the coherent
state\cite{QND1}. Therefor, the setup shown in Fig. 2 can achieve
the function of a PCM.

\section{ECP with the PCM}

With the PCM shown in Fig. 1, the principle of our ECP for
less-entangled W state is shown in Fig. 2. Three photons emitted
from the source S$_1$ are sent to Alice, Bob and Charlie from the
spatial modes $a1$, $b1$ and $c1$. The less-entangled photon pair is
described as
\begin{eqnarray}
|\Phi\rangle_{a1b1c1}&=&\alpha|V\rangle_{a1}|H\rangle_{b1}|H\rangle_{c1}+\beta|H\rangle_{a1}|V\rangle_{b1}|H\rangle_{c1}\nonumber\\
&+&\gamma|H\rangle_{a1}|H\rangle_{b1}|V\rangle_{c1},\label{W state}
\end{eqnarray}
 with
$|\alpha|^{2}+|\beta|^{2}+|\gamma|^{2}=1$. Then the single-photon
source S$_{2}$ emits two single photons to Alice and Charlie
respectively of the same form
\begin{eqnarray}
|\Phi\rangle_{a2}&=&|\Phi\rangle_{a3}=\frac{1}{\sqrt{2}}(|H\rangle+|V\rangle),\label{auxiliary1}
\end{eqnarray}
in the spatial mode $a2$ and $c2$, respectively. The five-photon
system
 can be written as
\begin{eqnarray}
|\Psi\rangle&=&|\Phi\rangle_{a1b1c1}\otimes|\Phi\rangle_{a2}\otimes|\Phi\rangle_{a3}
=(\alpha|V\rangle_{a1}|H\rangle_{b1}|H\rangle_{c1}\nonumber\\
&+&\beta|H\rangle_{a1}|V\rangle_{b1}|H\rangle_{c1}
+\gamma|H\rangle_{a1}|H\rangle_{b1}|V\rangle_{c1})\nonumber\\
&\otimes&(\frac{1}{\sqrt{2}}(|H\rangle_{a2}+|V\rangle_{a2}))\otimes(\frac{1}{\sqrt{2}}(|H\rangle_{c2}+|V\rangle_{c2}))\nonumber\\
&=&\alpha|V\rangle_{a1}|H\rangle_{b1}|H\rangle_{c1}\otimes
(|H\rangle_{a2}|H\rangle_{c2}+|H\rangle_{a2}|V\rangle_{c2}\nonumber\\
&+&|V\rangle_{a2}|H\rangle_{c2}+|V\rangle_{a2}|V\rangle_{c2})\nonumber\\
&+&\beta|H\rangle_{a1}|V\rangle_{b1}|H\rangle_{c1}\otimes(|H\rangle_{a2}|H\rangle_{c2}+|H\rangle_{a2}|V\rangle_{c2}\nonumber\\
&+&|V\rangle_{a2}|H\rangle_{c2}+|V\rangle_{a2}|V\rangle_{c2})\nonumber\\
&+&\gamma|H\rangle_{a1}|H\rangle_{b1}|V\rangle_{c1})\otimes(|H\rangle_{a2}|H\rangle_{c2}+|H\rangle_{a2}|V\rangle_{c2}\nonumber\\
&+&|V\rangle_{a2}|H\rangle_{c2}+|V\rangle_{a2}|V\rangle_{c2}).\label{combine1}
\end{eqnarray}

\begin{figure}[!h]
\begin{center}
\includegraphics[width=7cm,angle=0]{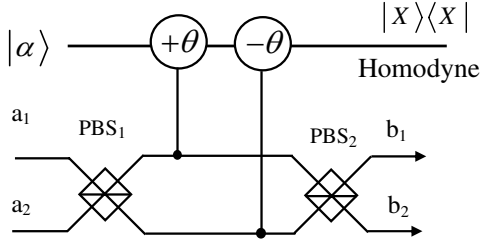}
\caption{The principle of a parity check measurement (PCM)
constructed with cross-Kerr nonlinearity, the same as that in
Refs.\cite{shengpra,zhangshou}. PBSs represent the polarizing beam
splitters which transmit horizontal polarization $|H\rangle$ and
reflect the vertical polarization $|V\rangle$. $\pm \theta$ denote
the different phase shift of the coherent state $|\alpha\rangle$
when there is a photon passing through the media. $|X\rangle\langle
X|$ represent an X quadrature measurement described in
Ref.\cite{QND1}.}
\end{center}
\end{figure}

Then Alice and Charlie both let his two photons pass through their
PCM gates respectively to make a parity check measurement. If they
both pick up the even parity states, the state of Eq.
(\ref{combine1}) collapses to
\begin{eqnarray}
|\Psi\rangle&\rightarrow&|\Psi'\rangle=\alpha|V\rangle_{a1}|H\rangle_{b1}|H\rangle_{c1}|V\rangle_{a2}|H\rangle_{c2}\nonumber\\
&+&\beta|H\rangle_{a1}|V\rangle_{b1}|H\rangle_{c1}|H\rangle_{a2}|H\rangle_{c2}\nonumber\\
&+&\gamma|H\rangle_{a1}|H\rangle_{b1}|V\rangle_{c1})|H\rangle_{a2}|V\rangle_{c2}\label{collapse1},
\end{eqnarray}
with a probability of 1/4. Certainly, there are other cases with the
same probability that Alice picks up the even parity states but
Charlie picks up the odd parity states, or Alice picks up the odd
parity states but Charlie picks up the even parity state. Also they
will both pick up the odd parity states
\begin{eqnarray}
&\rightarrow&\alpha|V\rangle_{a1}|H\rangle_{b1}|H\rangle_{c1}|H\rangle_{a2}|V\rangle_{c2}\nonumber\\
&+&\beta|H\rangle_{a1}|V\rangle_{b1}|H\rangle_{c1}|V\rangle_{a2}|V\rangle_{c2}\nonumber\\
&+&\gamma|H\rangle_{a1}|H\rangle_{b1}|V\rangle_{c1})|V\rangle_{a2}|H\rangle_{c2}.\label{collapse2}
\end{eqnarray}
 If Alice or Charlie picks up the odd parity states, he or she
performs a bit-flip operation $\sigma_{x}=|H\rangle\langle
V|+|V\rangle\langle H|$ on one of his or her photon to change it
into the both even parity state in Eq. (\ref{collapse1}). Then we
can rewrite the state $|\Psi'\rangle$ under the two orthogonal basis
\begin{eqnarray}
|\varphi_{1}\rangle_{a}=\frac{\alpha}{\sqrt{\alpha^{2}+\beta^{2}}}|H\rangle+\frac{\beta}{\sqrt{\alpha^{2}+\beta^{2}}}|V\rangle,\nonumber\\
|\varphi^{\perp}_{1}\rangle_{a}=\frac{\beta}{\sqrt{\alpha^{2}+\beta^{2}}}|H\rangle-\frac{\alpha}{\sqrt{\alpha^{2}+\beta^{2}}}|V\rangle,
\end{eqnarray}
and
\begin{eqnarray}
|\varphi_{2}\rangle_{c}=\frac{\gamma}{\sqrt{\gamma^{2}+\beta^{2}}}|H\rangle+\frac{\beta}{\sqrt{\gamma^{2}+\beta^{2}}}|V\rangle,\nonumber\\
|\varphi^{\perp}_{2}\rangle_{c}=\frac{\beta}{\sqrt{\gamma^{2}+\beta^{2}}}|H\rangle-\frac{\gamma}{\sqrt{\gamma^{2}+\beta^{2}}}|V\rangle,
\end{eqnarray}
i.e.,
\begin{eqnarray}
|\Psi'\rangle&=&\frac{\alpha\beta\gamma}{\sqrt{\alpha^{2}+\beta^{2}}\sqrt{\gamma^{2}+\beta^{2}}}(|V\rangle_{a1}|H\rangle_{b1}|H\rangle_{c1}\nonumber\\
&+&|H\rangle_{a1}|V\rangle_{b1}|H\rangle_{c1}+|H\rangle_{a1}|H\rangle_{b1}|V\rangle_{c1})|\varphi_{1}\rangle_{a}|\varphi_{2}\rangle_{c}\nonumber\\
&+&(\frac{\alpha\beta^{2}}{\sqrt{\alpha^{2}+\beta^{2}}\sqrt{\gamma^{2}+\beta^{2}}}|V\rangle_{a1}|H\rangle_{b1}|H\rangle_{c1}\nonumber\\
&+&\frac{\alpha\beta^{2}}{\sqrt{\alpha^{2}+\beta^{2}}\sqrt{\gamma^{2}+\beta^{2}}}|H\rangle_{a1}|V\rangle_{b1}|H\rangle_{c1}\nonumber\\
&-&\frac{\alpha\gamma^{2}}{\sqrt{\alpha^{2}+\beta^{2}}\sqrt{\gamma^{2}+\beta^{2}}}|H\rangle_{a1}|H\rangle_{b1}|V\rangle_{c1}|\varphi_{1}\rangle_{a}|\varphi^{\perp}_{2}\rangle_{c})\nonumber\\
&-&(\frac{\alpha^{2}\gamma}{\sqrt{\alpha^{2}+\beta^{2}}\sqrt{\gamma^{2}+\beta^{2}}}|V\rangle_{a1}|H\rangle_{b1}|H\rangle_{c1}\nonumber\\
&+&\frac{\beta^{2}\gamma}{\sqrt{\alpha^{2}+\beta^{2}}\sqrt{\gamma^{2}+\beta^{2}}}|H\rangle_{a1}|V\rangle_{b1}|H\rangle_{c1}\nonumber\\
&+&\frac{\beta^{2}\gamma}{\sqrt{\alpha^{2}+\beta^{2}}\sqrt{\gamma^{2}+\beta^{2}}}|H\rangle_{a1}|H\rangle_{b1}|V\rangle_{c1}|\varphi_{1}^{\perp}\rangle_{a}|\varphi_{2}\rangle_{c})\nonumber\\
&+&(\frac{\alpha^{2}\beta}{\sqrt{\alpha^{2}+\beta^{2}}\sqrt{\gamma^{2}+\beta^{2}}}|V\rangle_{a1}|H\rangle_{b1}|H\rangle_{c1}\nonumber\\
&+&\frac{\beta^{3}}{\sqrt{\alpha^{2}+\beta^{2}}\sqrt{\gamma^{2}+\beta^{2}}}|H\rangle_{a1}|V\rangle_{b1}|H\rangle_{c1}\nonumber\\
&+&\frac{\beta\gamma^{2}}{\sqrt{\alpha^{2}+\beta^{2}}\sqrt{\gamma^{2}
+\beta^{2}}}|H\rangle_{a1}|H\rangle_{b1}|V\rangle_{c1}|\varphi_{1}^{\perp}\rangle_{a}|\varphi_{2}^{\perp}\rangle_{c}).\label{deposion}\nonumber\\
\end{eqnarray}

\begin{figure}[!h]
\begin{center}
\includegraphics[width=9cm,angle=0]{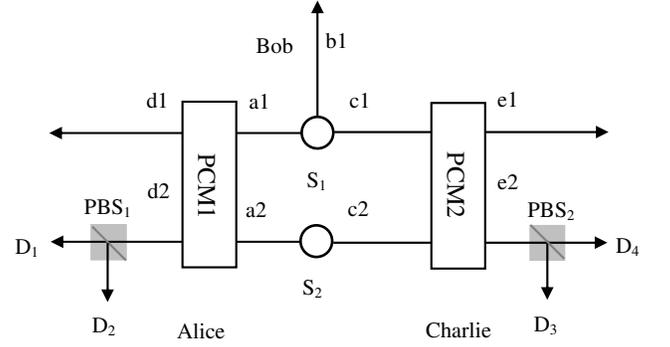}
\caption{ Schematic drawing of this ECP with two conventional single
photons of the form $\frac{1}{\sqrt{2}}(|H\rangle+|V\rangle)$.
S$_{1}$ is the less-entangled W state source and S$_{2}$ is the
single photon source. PCM1 and PCM2 are the same setup of parity
check measurement described in Fig. 1.}
\end{center}
\end{figure}
Alice uses  a PBS$_1$ whose optical axis is placed at the angle
$\varphi_{1}$, and two detectors D$_{1}$ and D$_{2}$ to complete the
measurement on the photon in the spatial mode $d2$ with the basis
$\{|\varphi_{1}\rangle_{a},|\varphi^{\perp}_{1}\rangle_{a}\}$. With
the same principle, Charlie uses  a PBS$_2$ whose optical axis is
placed at the angle $\varphi_{2}$, and two detectors D$_{3}$ and
D$_{4}$ to complete the measurement on the photon in the spatial
mode $d2$ with the basis
$\{|\varphi_{2}\rangle_{c},|\varphi^{\perp}_{2}\rangle_{c}\}$, shown
in Fig.2. Here $cos
\varphi_{1}=\frac{\alpha}{\sqrt{\alpha^{2}+\beta^{2}}} $, and  $sin
\varphi_{1}=-\frac{\beta}{\sqrt{\alpha^{2}+\beta^{2}}}$.  $cos
\varphi_{2}=\frac{\gamma}{\sqrt{\gamma^{2}+\beta^{2}}} $, and  $sin
\varphi_{1}=-\frac{-\beta}{\sqrt{\gamma^{2}+\beta^{2}}}$. From the
Eq. (\ref{deposion}), if Alice and Charlie obtain the states
$|\varphi_{1}\rangle_{a}$ and $|\varphi_{2}\rangle_{c}$,
respectively when they measure their photons in the spatial modes
$d2$ and $e2$, the remaining photon pair is essentially the
maximally entangled W state
$|\Phi\rangle_{a1b1c1}=\frac{1}{\sqrt{3}}(|V\rangle|H\rangle|H\rangle+|H\rangle|V\rangle|H\rangle+|H\rangle|H\rangle|V\rangle)$.
The success probability is
\begin{eqnarray}
P_{0}=\frac{3\alpha^{2}\beta^{2}\gamma^{2}}{(\alpha^{2}+\beta^{2})(\beta^{2}+\gamma^{2})}.
\end{eqnarray}
Otherwise, if Alice and Charlie obtain the other states, they cannot
get the maximally entangled state. For example, if Alice obtains
$|\varphi_{1}\rangle_{a}$ and Charlie obtains
$|\varphi^{\perp}_{2}\rangle_{c}$, they will get another
lesser-entangled state
\begin{eqnarray}
|\Phi_{1}\rangle_{a1b1c1}&=&(\frac{\alpha\beta^{2}}{\sqrt{\alpha^{2}+\beta^{2}}\sqrt{\gamma^{2}+\beta^{2}}}|V\rangle_{a1}|H\rangle_{b1}|H\rangle_{c1}\nonumber\\
&+&\frac{\alpha\beta^{2}}{\sqrt{\alpha^{2}+\beta^{2}}\sqrt{\gamma^{2}+\beta^{2}}}|H\rangle_{a1}|V\rangle_{b1}|H\rangle_{c1}\nonumber\\
&-&\frac{\alpha\gamma^{2}}{\sqrt{\alpha^{2}+\beta^{2}}\sqrt{\gamma^{2}+\beta^{2}}}|H\rangle_{a1}|H\rangle_{b1}|V\rangle_{c1},\nonumber\\
\end{eqnarray}
with the probability of
\begin{eqnarray}
P_{1}=\frac{2\alpha^{2}\beta^{4}+\alpha^{2}\gamma^{4}}{(\alpha^{2}+\beta^{2})(\beta^{2}+\gamma^{2})}.
\end{eqnarray}
 It can be written as
\begin{eqnarray}
|\Phi_{1}\rangle_{a1b1c1}&=&\frac{\beta^{2}}{\sqrt{\gamma^{4}+2\beta^{4}}}|V\rangle_{a1}|H\rangle_{b1}|H\rangle_{c1}\nonumber\\
&+&\frac{\beta^{2}}{\sqrt{\gamma^{4}+2\beta^{4}}}|H\rangle_{a1}|V\rangle_{b1}|H\rangle_{c1}\nonumber\\
&-&\frac{\gamma^{2}}{\sqrt{\gamma^{4}+2\beta^{4}}}|H\rangle_{a1}|H\rangle_{b1}|V\rangle_{c1}\label{less1}.
\end{eqnarray}
If Alice obtains the state $|\varphi^{\perp}_{1}\rangle_{a}$ and
Charlie obtains the state $|\varphi_{2}\rangle_{c}$, they will get
\begin{eqnarray}
|\Phi_{2}\rangle_{a1b1c1}&=&-\frac{\alpha^{2}}{\sqrt{\alpha^{4}+2\beta^{4}}}|V\rangle_{a1}|H\rangle_{b1}|H\rangle_{c1}\nonumber\\
&+&\frac{\beta^{2}}{\sqrt{\alpha^{4}+2\beta^{4}}}|H\rangle_{a1}|V\rangle_{b1}|H\rangle_{c1}\nonumber\\
&+&\frac{\beta^{2}}{\sqrt{\alpha^{4}+2\beta^{4}}}|H\rangle_{a1}|H\rangle_{b1}|V\rangle_{c1}\label{less2},
\end{eqnarray}
with the probability of
\begin{eqnarray}
P_{2}=\frac{\alpha^{4}\gamma^{2}+2\beta^{4}\gamma^{2}}{(\alpha^{2}+\beta^{2})(\beta^{2}+\gamma^{2})}.
\end{eqnarray}
If Alice obtain the state $|\varphi^{\perp}_{1}\rangle_{a}$ and
Charlie obtains the state $|\varphi^{\perp}_{2}\rangle_{c}$, they
will get
\begin{eqnarray}
|\Phi_{3}\rangle_{a1b1c1}&=&\frac{\alpha^{2}}{\sqrt{\alpha^{4}+\beta^{4}+\gamma^{4}}}|V\rangle_{a1}|H\rangle_{b1}|H\rangle_{c1}\nonumber\\
&+&\frac{\beta^{2}}{\sqrt{\alpha^{4}+\beta^{4}+\gamma^{4}}}|H\rangle_{a1}|V\rangle_{b1}|H\rangle_{c1}\nonumber\\
&+&\frac{\gamma^{2}}{\sqrt{\alpha^{4}+\beta^{4}+\gamma^{4}}}|H\rangle_{a1}|H\rangle_{b1}|V\rangle_{c1}\label{less3},
\end{eqnarray}
with the probability of
\begin{eqnarray}
P_{3}=\frac{\alpha^{4}\beta^{2}+\beta^{6}+\beta^{2}\gamma^{4}}{(\alpha^{2}+\beta^{2})(\beta^{2}+\gamma^{2})}.
\end{eqnarray}
\begin{figure}[!h]
\begin{center}
\includegraphics[width=6cm,angle=0]{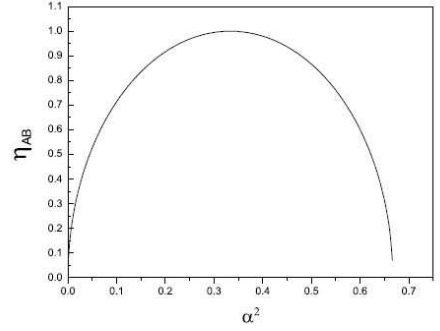}
\caption{The relationship between the entanglement transformation
efficiency $\eta_{AB}$ of the present ECP and the initial
coefficient of the less-entangled state $\alpha^{2}$, when
$\gamma=\frac{1}{\sqrt{3}}$. We change
$\alpha\in(0,\sqrt{\frac{2}{3}})$. $\eta_{AB}$ can reach 1 when
$\alpha=\beta=\gamma=\frac{1}{\sqrt{3}}$.}
\end{center}
\end{figure}

Interestingly, all states of Eqs. (\ref{less1}), (\ref{less2}) and
(\ref{less3}) are the lesser-entangled W states. We first compare
the state of Eq. (\ref{less3}) with the original less-entangled
state of Eq. (\ref{W state}). They have essentially the same form.
That is, we  choose
$\alpha'=\frac{\alpha^{2}}{\sqrt{\alpha^{4}+\beta^{4}+\gamma^{4}}}$,
$\beta'=\frac{\beta^{2}}{\sqrt{\alpha^{4}+\beta^{4}+\gamma^{4}}}$
and
$\gamma'=\frac{\gamma^{2}}{\sqrt{\alpha^{4}+\beta^{4}+\gamma^{4}}}$.
In order to obtain the maximally entangled W state from Eq.
(\ref{less3}), they only need to perform the ECP in a second round,
following the same principle described above.

Now let us discuss another states of Eqs. (\ref{less1}) and
(\ref{less2}). Both $|\Phi_{1}\rangle_{a1b1c1}$ and
$|\Phi_{2}\rangle_{a1b1c1}$ are the lesser-entangled states, but
they are different from the $|\Phi_{3}\rangle_{a1b1c1}$, because
they  essentially only have two different coefficients. They also
can  be concentrated into a maximally entangled state.
Interestingly, this kind of states are much easier to be
concentrated than the original state $|\Phi\rangle_{a1b1c1}$,
because one of them only need one single-photon to perform this ECP.
We take state of Eq. (\ref{less1}) as an example. If they obtain Eq.
(\ref{less1}), they first perform a phase-flip operation and make it
become
\begin{eqnarray}
|\Phi_{1}'\rangle_{a1b1c1}&=&\frac{\beta^{2}}{\sqrt{\gamma^{4}+2\beta^{4}}}|V\rangle_{a1}|H\rangle_{b1}|H\rangle_{c1}\nonumber\\
&+&\frac{\beta^{2}}{\sqrt{\gamma^{4}+2\beta^{4}}}|H\rangle_{a1}|V\rangle_{b1}|H\rangle_{c1}\nonumber\\
&+&\frac{\gamma^{2}}{\sqrt{\gamma^{4}+2\beta^{4}}}|H\rangle_{a1}|H\rangle_{b1}|V\rangle_{c1}.\label{less4}
\end{eqnarray}

Then source of S$_2$ emits a single photon to Charlie, and the
four-photon state can be written as
\begin{eqnarray}
&&|\Phi_{1}'\rangle_{a1b1c1}\otimes\frac{1}{\sqrt{2}}(|H\rangle_{c2}+|V\rangle_{c2})\nonumber\\
&=&(\frac{\beta^{2}}{\sqrt{\gamma^{4}+2\beta^{4}}}|V\rangle_{a1}|H\rangle_{b1}|H\rangle_{c1}\nonumber\\
&+&\frac{\beta^{2}}{\sqrt{\gamma^{4}+2\beta^{4}}}|H\rangle_{a1}|V\rangle_{b1}|H\rangle_{c1}\nonumber\\
&+&\frac{\gamma^{2}}{\sqrt{\gamma^{4}+2\beta^{4}}}|H\rangle_{a1}|H\rangle_{b1}|V\rangle_{c1})\otimes\frac{1}{\sqrt{2}}(|H\rangle_{c2}+|V\rangle_{c2})\nonumber\\
&=&\frac{1}{\sqrt{2}}(\frac{\beta^{2}}{\sqrt{\gamma^{4}+2\beta^{4}}}|V\rangle_{a1}|H\rangle_{b1}|H\rangle_{c1}|H\rangle_{c2}\nonumber\\
&+&\frac{\beta^{2}}{\sqrt{\gamma^{4}+2\beta^{4}}}|H\rangle_{a1}|V\rangle_{b1}|H\rangle_{c1}|H\rangle_{c2}\nonumber\\
&+&\frac{\gamma^{2}}{\sqrt{\gamma^{4}+2\beta^{4}}}|H\rangle_{a1}|H\rangle_{b1}|V\rangle_{c1}|V\rangle_{c2})\nonumber\\
&+&\frac{1}{\sqrt{2}}(\frac{\beta^{2}}{\sqrt{\gamma^{4}+2\beta^{4}}}|V\rangle_{a1}|H\rangle_{b1}|H\rangle_{c1}|V\rangle_{c2}\nonumber\\
&+&\frac{\beta^{2}}{\sqrt{\gamma^{4}+2\beta^{4}}}|H\rangle_{a1}|V\rangle_{b1}|H\rangle_{c1}|V\rangle_{c2}\nonumber\\
&+&\frac{\gamma^{2}}{\sqrt{\gamma^{4}+2\beta^{4}}}|H\rangle_{a1}|H\rangle_{b1}|V\rangle_{c1}|H\rangle_{c2}).\label{combine3}
\end{eqnarray}
After passing through the PCM, if Charlie picks up the even parity
state, Eq. (\ref{combine3}) collapses to
\begin{eqnarray}
&\rightarrow&\frac{\beta^{2}}{\sqrt{\gamma^{4}+2\beta^{4}}}|V\rangle_{a1}|H\rangle_{b1}|H\rangle_{c1}|H\rangle_{c2}\nonumber\\
&+&\frac{\beta^{2}}{\sqrt{\gamma^{4}+2\beta^{4}}}|H\rangle_{a1}|V\rangle_{b1}|H\rangle_{c1}|H\rangle_{c2}\nonumber\\
&+&\frac{\gamma^{2}}{\sqrt{\gamma^{4}+2\beta^{4}}}|H\rangle_{a1}|H\rangle_{b1}|V\rangle_{c1}|V\rangle_{c2}.
\end{eqnarray}
We can rewrite above state under the orthogonal basis
\begin{eqnarray}
|\varphi_{2}\rangle'_{c}=\frac{\gamma^{2}}{\sqrt{\gamma^{4}+\beta^{4}}}|H\rangle+\frac{\beta^{2}}{\sqrt{\gamma^{4}+\beta^{4}}}|V\rangle,\nonumber\\
|\varphi^{\perp}_{2}\rangle'_{c}=\frac{\beta^{2}}{\sqrt{\gamma^{4}+\beta^{4}}}|H\rangle-\frac{\gamma^{2}}{\sqrt{\gamma^{4}+\beta^{4}}}|V\rangle.
\end{eqnarray}

\begin{figure}[!h]
\begin{center}
\includegraphics[width=6cm,angle=0]{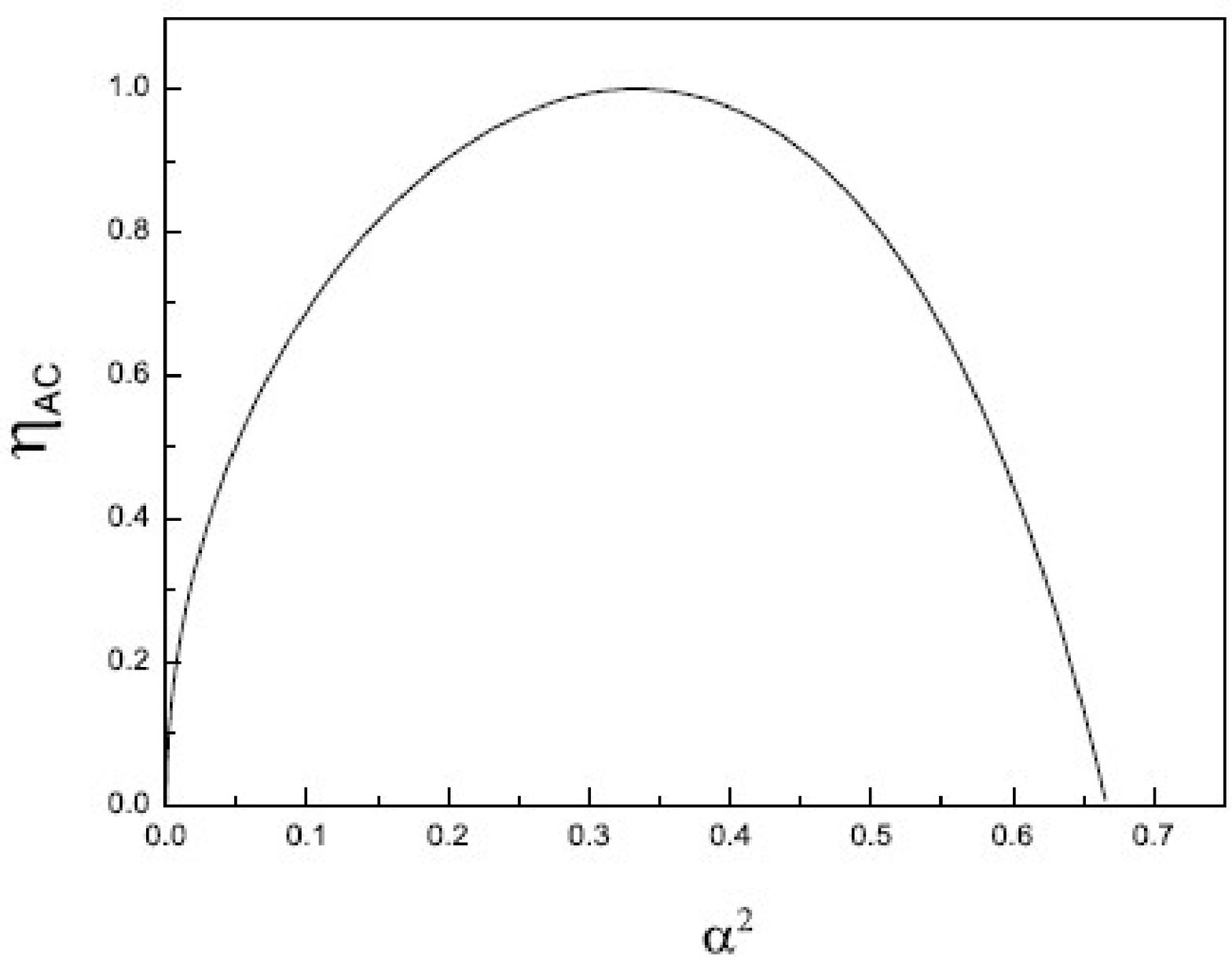}
\caption{The relationship between the entanglement transformation
efficiency $\eta_{AC}$ of the present ECP and the initial
coefficient of the less-entangled state $\alpha^{2}$, when
$\gamma=\frac{1}{\sqrt{3}}$. We change
$\alpha\in(0,\sqrt{\frac{2}{3}})$.}
\end{center}
\end{figure}

Then Charlie lets his photon in mode $e2$ pass through the
PBS$_{2}$, whose optical axis is placed at the angle $\varphi'_{2}$,
and two detectors to complete the measurement on the photon in the
mode $e2$. Here
$cos\varphi'_{2}=\frac{\gamma^{2}}{\sqrt{\gamma^{4}+\beta^{4}}}$ and
$sin \varphi'_{2}=-\frac{\beta^{2}}{\sqrt{\gamma^{4}+\beta^{4}}}$.
 If Charlie obtains the state $|\varphi_{2}\rangle'_{c}$, they will
 obtain the maximally entangled state. If Charlie obtains
 $|\varphi^{\perp}_{2}\rangle'_{c}$, they will obtain another
 lesser-entangled state of the form
\begin{eqnarray}
&\rightarrow&\frac{\beta^{4}}{\sqrt{\gamma^{4}+2\beta^{4}}\sqrt{\gamma^{4}+\beta^{4}}}|V\rangle_{a1}|H\rangle_{b1}|H\rangle_{c1}\nonumber\\
&+&\frac{\beta^{4}}{\sqrt{\gamma^{4}+2\beta^{4}}\sqrt{\gamma^{4}+\beta^{4}}}|H\rangle_{a1}|V\rangle_{b1}|H\rangle_{c1}\nonumber\\
&-&\frac{\gamma^{4}}{\sqrt{\gamma^{4}+2\beta^{4}}\sqrt{\gamma^{4}+\beta^{4}}}|H\rangle_{a1}|H\rangle_{b1}|V\rangle_{c1}.\label{less4}
\end{eqnarray}
Eq. (\ref{less4}) is also a lesser-entangled W state which can be
reconcentrated into a maximally entangled W state in a third round.
Certainly, from Eq. (\ref{combine3}), if Charlie obtains an odd
parity state, he needs  to perform a bit-flip operation. Then the
following steps are the same as described above. In this way, they
can ultimately obtain a maximally entangled W state by repeating
this ECP. On the other hand, if the measurements results of the
first time performed by Alice and Charlie are
$|\varphi^{\perp}_{1}\rangle_{a}$ and $|\varphi_{2}\rangle_{c}$, and
lead to the collapsed state of Eq. (\ref{less2}), it can also be
reconcentrated in a second round and only Alice needs to perform the
ECP similar to Charlie's operation described above. That is, all the
remaining lesser-entangled states can be reconcentrated. If the
remaining state is Eq. (\ref{less1}), only Charlie needs to repeat
this ECP. If the remaining state is  Eq. (\ref{less2}), only Alice
needs to repeat this ECP, and if the remaining state is Eq.
(\ref{less3}), both Alice and Charlie should repeat the whole ECP,
following the same principle described above.

\section{entanglement transformation efficiency}

Thus far, we have fully described this ECP. It is well known that
LOCC cannot increase entanglement. Therefore, this ECP under LOCC
must be the entanglement transformation. Similar to Ref.
\cite{shengpra3}, we can calculate the entanglement transformation
efficiency after performing this ECP. In Ref. \cite{shengpra3}, we
calculated the entanglement transformation efficiency for GHZ state
concentration. However, different from the GHZ state, the W state
has the different entanglement structure from the GHZ state. The
entanglement of GHZ state is a global entanglement. That is, if we
trace over any one of the particle, the remaining particles do not
entangle anymore. But the W state is the entanglement between each
particles. If we trace over any one of the particle, the remaining
two particles still entangle. Therefore, for a W state described in
Eq. (\ref{W state}), the entanglement of three-tangle is quale to 0
\cite{coffman,plesch,baokui1,baokui2}.

In this paper, we use the concurrence described in Refs.
\cite{concurrrence,coffman} to calculate the entanglement between
each particles.
\begin{figure}[!h]
\begin{center}
\includegraphics[width=6cm,angle=0]{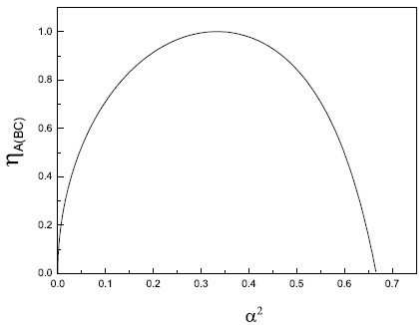}
\caption{The relationship between the entanglement transformation
efficiency $\eta_{A(BC)}$ of the present ECP and the initial
coefficient of the less-entangled state $\alpha^{2}$, when
$\gamma=\frac{1}{\sqrt{3}}$. We change
$\alpha\in(0,\sqrt{\frac{2}{3}})$.}
\end{center}
\end{figure}
\begin{figure}[!h]
\begin{center}
\includegraphics[width=6cm,angle=0]{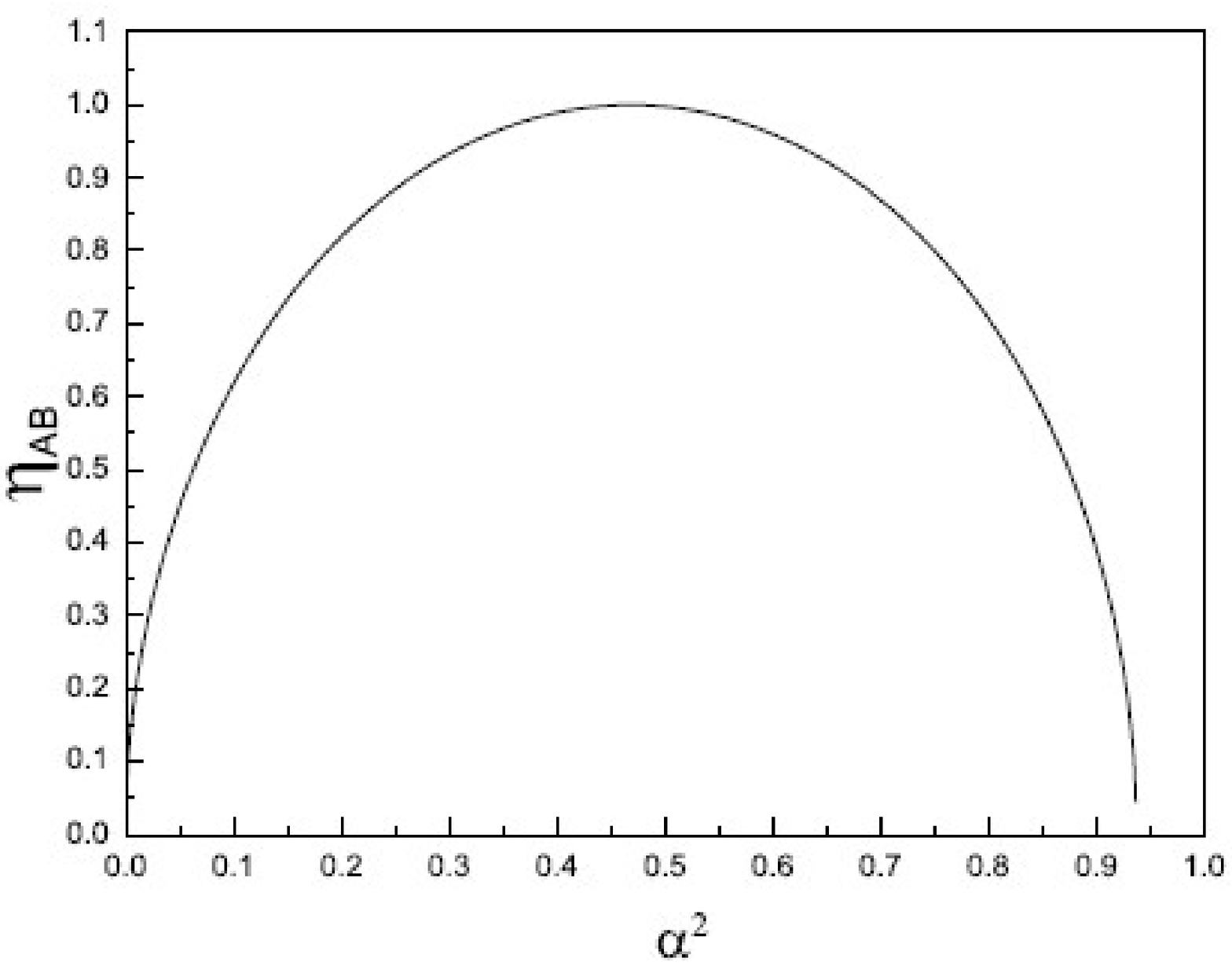}
\caption{The relationship between the entanglement transformation
efficiency $\eta_{AB}$ of the present ECP and the initial
coefficient of the less-entangled state $\alpha^{2}$, when
$\gamma=\frac{1}{4}$. We change
$\alpha\in(0,\sqrt{\frac{15}{32}})$.}
\end{center}
\end{figure}
\begin{figure}[!h]
\begin{center}
\includegraphics[width=6cm,angle=0]{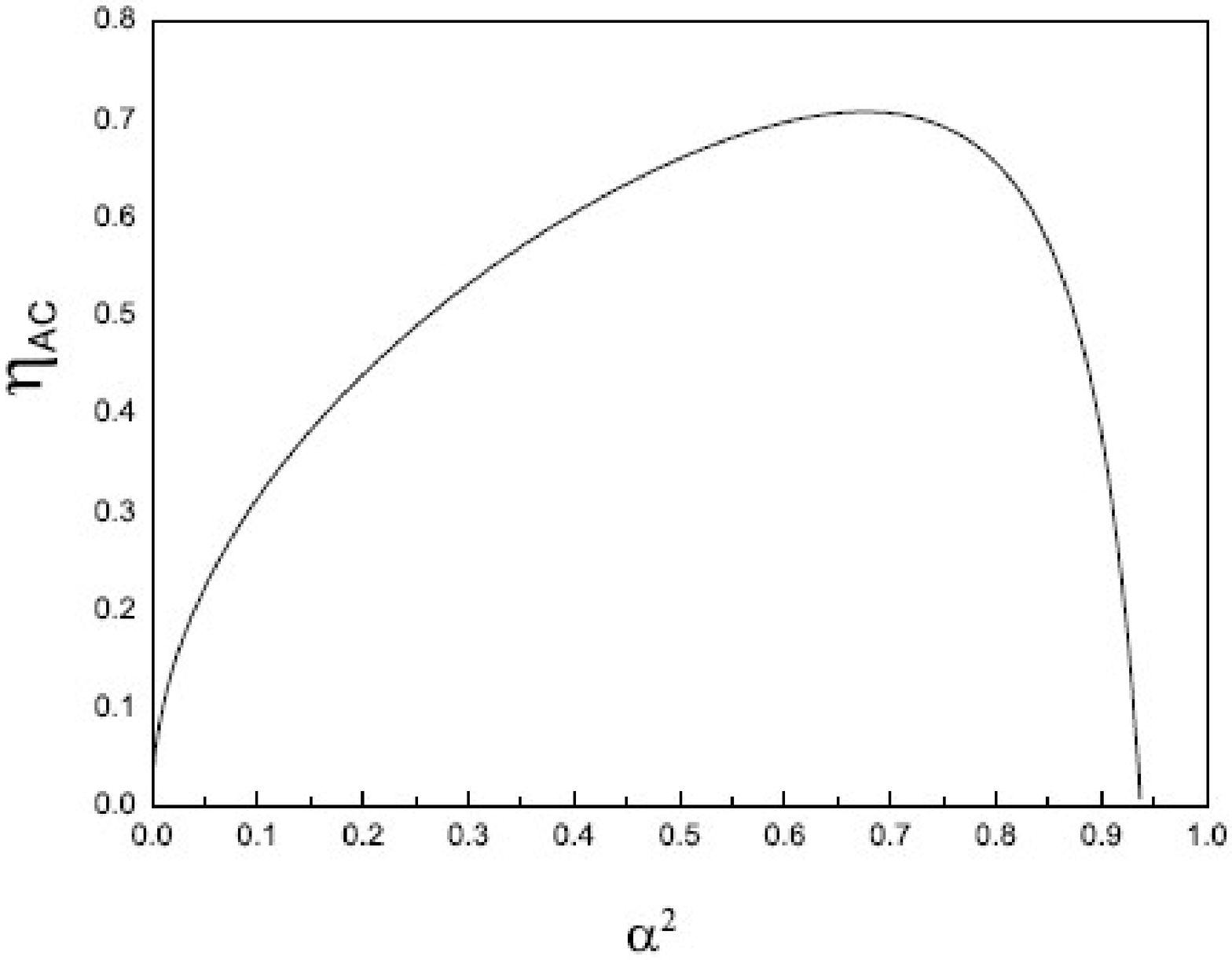}
\caption{The relationship between the entanglement transformation
efficiency $\eta_{AC}$ of the present ECP and the initial
coefficient of the less-entangled state $\alpha^{2}$, when
$\gamma=\frac{1}{4}$. We change
$\alpha\in(0,\sqrt{\frac{15}{32}})$.}
\end{center}
\end{figure}

\begin{figure}[!h]
\begin{center}
\includegraphics[width=6cm,angle=0]{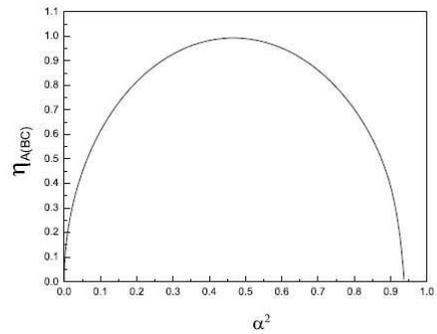}
\caption{The relationship between the entanglement transformation
efficiency $\eta_{A(BC)}$ of the present ECP and the initial
coefficient of the less-entangled state $\alpha^{2}$, when
$\gamma=\frac{1}{4}$. We change
$\alpha\in(0,\sqrt{\frac{15}{32}})$.}
\end{center}
\end{figure}

Before performing the ECP, the concurrence between particles A and B
in Eq. (\ref{W state}) is $C_{AB}=2|\alpha\beta|$. The concurrence
between particles A and C in Eq. (\ref{W state}) is
$C_{AC}=2|\alpha\gamma|$, and the concurrence of the two subsystem A
and BC is $C_{A(BC)}=2|\alpha|\sqrt{|\beta|^{2}+|\gamma|^{2}}$. We
denote the entanglement transformation efficiency as
\begin{eqnarray}
\eta_{AB}=\frac{C'_{AB}}{C_{AB}},\nonumber\\
 \eta_{AC}=\frac{C'_{AC}}{C_{AC}},\nonumber\\
 \eta_{A(BC)}=\frac{C'_{A(BC)}}{C_{A(BC)}}.
\end{eqnarray}
Here
\begin{eqnarray}
{C'_{AB}}&=&\frac{2}{3}P_{0}+\frac{2\beta^{4}}{\gamma^{4}+2\beta^{4}}P_{1}+\frac{2\alpha^{2}\beta^{2}}{\alpha^{4}+2\beta^{4}}P_{2}\nonumber\\
&+&\frac{2\alpha^{2}\beta^{2}}{\alpha^{4}+\beta^{4}+\gamma^{4}}P_{3}.\label{Cab}
\end{eqnarray}
The $C'_{AB}$ is the concurrence between particle A and B after
performing this ECP. From Eq. (\ref{Cab}), it comprises four terms.
The first term $\frac{2}{3}P_{0}$ means that after concentration,
they obtain the maximally entangled W state with the probability of
$P_0$ and it describes the concurrence between A and B for the
maximally entangled W state. The second term and other terms are
analogy with the first one. For instance, the second term means that
with the probability of $P_1$, they obtain the state of Eq.
(\ref{less1}), and item $\frac{2\beta^{4}}{\gamma^{4}+2\beta^{4}}$
describes the concurrence between A and B if they obtain such state.
$C'_{AC}$ and $C_{A(BC)}$ are analogy with $C'_{AB}$. $C'_{AC}$
means the concurrence between particles A and C after concentration
and $C'_{A(BC)}$ means the concurrence between subsystem A and BC
after performing the ECP.
\begin{eqnarray}
C'_{AC}&=&\frac{2}{3}P_{0}+\frac{2\beta^{2}\gamma^{2}}{\gamma^{4}+2\beta^{4}}P_{1}+\frac{2\alpha^{2}\beta^{2}}{\alpha^{4}+2\beta^{4}}P_{2}\nonumber\\
&+&\frac{2\alpha^{2}\gamma^{2}}{\alpha^{4}+\beta^{4}+\gamma^{4}}P_{3}.
\end{eqnarray}
\begin{eqnarray}
&&C'_{A(BC)}=\frac{2\sqrt{2}}{3}P_{0}\nonumber\\
&+&\frac{2\beta^{2}}{\sqrt{\gamma^{4}+2\beta^{4}}}\sqrt{\frac{\beta^{4}}{\gamma^{4}+2\beta^{4}}+\frac{\beta^{4}}{\gamma^{4}+2\beta^{4}}}P_{1}\nonumber\\
&+&\frac{2\alpha^{2}}{\sqrt{\alpha^{4}+2\beta^{4}}}\sqrt{\frac{\beta^{4}}{\alpha^{4}+2\beta^{4}}+\frac{\beta^{4}}{\alpha^{4}+2\beta^{4}}}P_{2}\nonumber\\
&+&\frac{2\alpha^{2}}{\sqrt{\alpha^{4}+\beta^{4}+\gamma^{4}}}\sqrt{\frac{\beta^{4}}{\alpha^{4}+\beta^{4}+\gamma^{4}}+\frac{\gamma^{4}}{\alpha^{4}+\beta^{4}+\gamma^{4}}}P_{3}.\nonumber\\
\end{eqnarray}
We calculate the entanglement transformation efficiency $\eta_{AB}$,
$\eta_{AC}$ and $\eta_{A(BC)}$ altered by the coefficient $\alpha$.
Here, we choose $\gamma=\frac{1}{\sqrt{3}}$, change
$\alpha\in(0,\sqrt{\frac{2}{3}})$. The relationship between the
coefficient $\alpha^{2}$ and entanglement transformation efficiency
$\eta_{AB}$ is shown in Fig. 3. Fig. 4 and Fig. 5 show the
relationship between $\alpha^{2}$ and $\eta_{AC}$ and
$\eta_{A(BC)}$, respectively. It is interesting to see that if
$\alpha=\beta=\gamma=\frac{1}{\sqrt{3}}$, the
$\eta_{AB}=\eta_{AC}=\eta_{A(BC)}=1$. The $\eta_{AB}$, $\eta_{AC}$,
and $\eta_{A(BC)}$ are not fixed values but all increases with the
initial entanglement. They can reach the maximally value 1 when
$\alpha=\frac{1}{\sqrt{3}}$. These result are consistent with the
result shown in Ref.\cite{shengpra3}. Interestingly, we also
calculate the entanglement transformation efficiency $\eta_{AB}$,
$\eta_{AC}$ and $\eta_{A(BC)}$ altered by the coefficient $\alpha$
when $\gamma=\frac{1}{4}$. In this case, the original state cannot
reach the maximally entangled state. From Fig. 6, $\eta_{AB}$ can
also reach the max value 1 when
$\alpha^{2}=\beta^{2}=\frac{1-(\frac{1}{4})^{2}}{2}=\frac{15}{32}$.
But $\eta_{AC}$ cannot reach 1 when we change $\alpha$, shown in
Fig. 7. Our numerical simulation shows that the max value
$\eta_{AC}\approx0.7078$ when $\alpha\approx0.6757$. In Fig. 8, we
calculate the $\eta_{A(BC)}$ alters with $\alpha^{2}$. It is shown
that $\eta_{A(BC)}=1$ when $\alpha^{2}=\frac{15}{32}$, which is
similar to Fig. 6.

\section{discussion and summary}
Thus far, we have fully explained this ECP. In this paper, the most
important element for us to complete this ECP is the PCM shown in
Fig. 1, constructed by cross-Kerr nonlinearity. At present, in the
optical single photon regime, it is still a quite-controversial
assumption to have a clean cross-Kerr nonlinearity. Natural
cross-Kerr nonlinearity is only $\tau\approx10^{-18}$ in the optical
single-photon regime\cite{kok1,kok2}. On the other hand, Hofmann et
al. pointed out that with a single two-level atom in a one-side
cavity, a $\pi$ phase shift can be reached\cite{hofmann}.
Fortunately, the PCM in our ECP does not require a strong
nonlinearity and it works for weak vales of the cross-Kerr coupling.
Gea-Banacloche argued that it is impossible to obtain a giant Kerr
effect with a single-photon wave packet\cite{Banacloche}, which is
consistent with the Shapiro and Razavi\cite{Shapiro1,Shapiro2}. He
\emph{et al.} showed that the high fidelities, nonzero conditional
phases and high photon numbers are compatible if the transverse-mode
effects can be suppressed\cite{he3}. Feizpour\emph{ et al.} also
showed that it is possible to amplify a cross-Kerr phase shift to an
observable value\cite{weak_meaurement}. Recently, Zhu and Zhang
discussed that coupled with a four-level, double $\Lambda$-type
configuration, giant cross-Kerr nonlinearities was also obtained
with nearly vanishing optical absorption\cite{oe}.

In summary, we present an efficient ECP for an arbitrary
three-photon W state. Alice and Charlie exploit the same PCM based
on the cross-Kerr nonlinearity to perform this ECP. By iterating
this protocol several times, it can reach a higher success
probability.  This ECP is also quite different from the conventional
ECPs for W state. First, it is focused on an arbitrary W state and
does not resort to the unitary evolution.  Second, it only requires
single photons of the same form $\frac{1}{2}(|H\rangle+|V\rangle)$.
Third, this PCM can works on the weak regime of cross-Kerr
nonlinearity, which is suitable for current technology.  We hope
this ECP maybe useful in a practical quantum information processing.

\section*{ACKNOWLEDGEMENTS}
We thank Dr. Bao-Kui Zhao (Jilin University) for helpful discussion.
This work was supported by the National Natural Science Foundation
of China under Grant No. 11104159, the Scientific Research
Foundation of Nanjing University of Posts and Telecommunications
under Grant No. NY211008, the University Natural Science Research
Foundation of JiangSu Province under Grant No. 11KJA510002, and the
open research fund of the Key Lab of Broadband Wireless
Communication and Sensor Network Technology (Nanjing University of
Posts and Telecommunications), Ministry of Education, China, and a
Project Funded by the Priority Academic Program Development of
Jiangsu Higher Education Institutions.


\begin{thebibliography}{99}

\bibitem{computation1} D. P. Divincenzo, Science  \textbf{270}, 255 (1995).

\bibitem{computation2} C. H. Bennett and D. P. Divincenzo,
Nature (London) \textbf{404}, 247 (2000).

\bibitem{rmp} N. Gisin, G. Ribordy, W. Tittel, and H. Zbinden,
Rev. Mod. Phys.  \textbf{74}, 145  (2002).

\bibitem{teleportation} C. H. Bennett, G. Brassard, C. Crepeau, R. Jozsa, A. Peres, and W. K. Wootters, Phys. Rev. Lett.
\textbf{70}, 1895 (1993).

\bibitem{cteleportation} A. Karlsson, M. Bourennane,
Phys. Rev. A \textbf{58}, 4394 (1998); F. G. Deng, C. Y. Li, Y. S.
Li, H. Y. Zhou, and Y. Wang, Phys. Rev. A \textbf{72}, 022338
(2005).

\bibitem{densecoding} C. H. Bennett and S. J. Wiesner, Phys. Rev. Lett. \textbf{69}, 2881 (1992).

\bibitem{Ekert91} A. K. Ekert,  Phys. Rev. Lett. \textbf{67},
  661 (1991).

  \bibitem{QSS1} M. Hillery, V. Bu\v{z}ek, and A. Berthiaume, Phys.
Rev. A \textbf{59}, 1829(1999).

\bibitem{QSS2} A. Karlsson, M. Koashi, and N. Imoto, Phys. Rev. A, \textbf{59},
162 (1999).

\bibitem{QSS3} L. Xiao, G.-L. Long, F.-G. Deng, and J.-W. Pan, Phys.
Rev. A \textbf{69}, 052307 (2004).



\bibitem{QSDC1} G.-L. Long, and X.-S. Liu, Phys. Rev. A, \textbf{65},
032302 (2002).

\bibitem{QSDC2} F.-G. Deng, G.-L. Long, and X.-S. Liu, Phys. Rev. A, \textbf{68},
042317 (2003).

\bibitem{QSDC3} C. Wang, F. G. Deng, Y. S. Li, X. S. Liu, and  G. L. Long,
Phys. Rev. A \textbf{71}, 042305 (2005).


\bibitem{QSTS1} R. Cleve, D. Gottesman,  H.K. Lo,
Phys. Rev. Lett. \textbf{83}  648, (1999) .

\bibitem{QSTS2} A.M. Lance, T. Symul, W.P. Bowen, B.C. Sanders,  P.K. Lam, Phys. Rev. Lett.  \textbf{92}
177903, (2004).

\bibitem{QSTS3} F. G. Deng, X. H. Li, C. Y. Li, P. Zhou, and H. Y.
Zhou, Phys. Rev. A \textbf{72}, 044301 (2005); F. G. Deng, X. H. Li,
C. Y. Li, P. Zhou, and H. Y. Zhou, Europ. Phys. J. D \textbf{39},
459 (2006); X. H. Li, P. Zhou, C. Y. Li, H. Y. Zhou, and F. G. Deng,
J. Phys. B \textbf{39}, 1975 (2006).





\bibitem{C.H.Bennett1} C. H. Bennett, G. Brassard, S. Popescu, B. Schumacher, J. A. Smolin and W. K. Wootters, Phys. Rev. Lett. \textbf{76}, 722 (1996).

\bibitem{D. Deutsch} D. Deutsch, A. Ekert, R. Jozsa, C. Macchiavello, S. Popescu and A. Sanpera, Phys. Rev. Lett. \textbf{77}, 2818 (1996).

\bibitem{M. Murao} M. Murao, M. B. Plenio, S. Popescu, V. Vedral and P. L. Knight, Phys. Rev. A \textbf{57}, R4075 (1998).

\bibitem{M. Horodecki} M. Horodecki and P. Horodecki, Phys. Rev. A \textbf{59}, 4206 (1999).

\bibitem{Pan1} J. W. Pan, C. Simon, and A. Zellinger,  Nature
\textbf{410}, 1067 (2001).
\bibitem{Simon} C. Simon  and  J. W. Pan,
Phys. Rev. Lett. \textbf{89}, 257901 (2002).

\bibitem{Pan2} J. W. Pan, S. Gasparonl, R. Ursin, G. Weihs and A. Zellinger, Nature   \textbf{423}, 417
(2003).
\bibitem{Yong} Y. W. Cheong, S. W. Lee, J. Lee and H. W. Lee, Phys. Rev. A \textbf{76}, 042314 (2007).

\bibitem{sangouard1} N. Sangouard, C. Simon, T. Coudreau, and N.
Gisin,
 Phys. Rev. A  \textbf{78}, 050301(R) (2008).

 \bibitem{wangc1} L. Xiao, C. Wang, W. Zhang, Y. D. Huang, J. D. Peng  and
G. L. Long,   Phys. Rev. A  \textbf{77}, 042315 (2008).

 \bibitem{sangouard2}D. Salart, O. Landry, N. Sangouard,
 N. Gisin, H. Herrmann, B. Sanguinetti, C. Simon, W. Sohler, R. T. Thew, A. Thomas, and H.
 Zbinden, Phys. Rev. Lett. \textbf{104}, 180504(2010).

\bibitem{shengpra} Y. B. Sheng, F. G. Deng, and H. Y. Zhou, Phys.
Rev. A \textbf{77}, 042308 (2008); Y. B. Sheng, and F. G. Deng,
\emph{ibid}. \textbf{81}, 032307 (2010); Y. B. Sheng, and F. G.
Deng, \emph{ibid}. \textbf{82}, 044305 (2010); Y. B. Sheng, F. G.
Deng, and H. Y. Zhou, Europ. Phys. J. D \textbf{55}, 235 (2009); Y.
B. Sheng, F. G. Deng, and G. L. Long, Phys. Lett. A  \textbf{375},
396 (2011).


\bibitem{lixhepp} X.  H. Li,  Phys.
Rev. A  \textbf{82},  044304 (2010).



\bibitem{wangc2} C. Wang, Y. Zhang  and G. S. Jin,
Phys. Rev. A \textbf{84},  032307 ( 2011).

\bibitem{wangc3} C. Wang, Y. Zhang  and G. S. Jin, Quant. Inf. Comput.
\textbf{11}, 988 (2011).




\bibitem{dengonestep} F.  G. Deng,   Phys. Rev. A \textbf{83},  062316
(2011); F.  G.  Deng, \emph{ibid}, \textbf{84}, 052312 (2011).





\bibitem{C.H.Bennett2} C. H. Bennett, H. J. Bernstein, S. Popescu, and
B. Schumacher, Phys. Rev. A \textbf{53}, 2046 (1996).

\bibitem{swapping1} S. Bose, V. Vedral, and P. L. Knight, Phys. Rev A
\textbf{60}, 194 (1999).
\bibitem{swapping2} B. S. Shi, Y. K. Jiang, and G. C. Guo, Phys.
Rev. A \textbf{62}, 054301 (2000).

\bibitem{bose}N. Paunkovi\'{c}, Y. Omar, S. Bose, and V. Vedral,
Phys. Rev. Lett. \textbf{88}, 187903 (2002).

\bibitem{zhao1} Z. Zhao, J. W. Pan, and M. S. Zhan, Phys. Rev. A \textbf{64}, 014301 (2001).

\bibitem{zhao2} Z. Zhao, T. Yang, Y. A. Chen, A. N. Zhang, and J. W.
Pan, Phys. Rev. Lett. \textbf{90}, 207901 (2003).

\bibitem{Yamamoto1} T. Yamamoto, M. Koashi, and N. Imoto, Phys. Rev.
A \textbf{64}, 012304 (2001).

\bibitem{Yamamoto2}T. Yamamoto, M. Koashi, S. K. Ozdemir, and N.
Imoto, Nature \textbf{421 }343 (2003).




\bibitem{wangxb} W. Xiang-bin, and F. Heng, Phys. Rev. A \textbf{68},
060302 (2003).

\bibitem{cao1}Z. L. Cao, L. H. Zhang, and M. Yang, Phys. Rev. A
\textbf{71,} 044302 (2005).

\bibitem{cao2}M. Yang, Y. Zhao, W. Song, and Z. L. Cao, Phys. Rev. A
\textbf{71}, 044302(2005).


\bibitem{shengpra2} Y. B. Sheng, F. G. Deng, and H. Y. Zhou, Phys.
Rev. A \textbf{77}, 062325 (2008).

\bibitem{shengpra3}Y. B. Sheng, L. Zhou, S. M. Zhao, and B. Y.
Zheng, Phys. Rev. A \textbf{85}, 012307 (2012).

\bibitem{shengpla}  Y. B. Sheng, F. G. Deng, and H. Y. Zhou, Phys.
Lett. A \textbf{373}, 1823 (2009).

\bibitem{shengqic}  Y. B. Sheng, F. G. Deng, and H. Y. Zhou, Quant.
Inf.  Comput. \textbf{10}, 272 (2010).

\bibitem{dengconcentration} F. G. Deng, Phys. Rev. A \textbf{85}, 022311
(2012).

\bibitem{zhang} C. W. Zhang, Quant. Inf.  Comput. \textbf{4}, 196 (2004).





\bibitem{dur} W. Dur, G. Vidal, and J. I. Cirac, Phys. Rev. A \textbf{62},
062314 (2000).




\bibitem{cao} Z. L. Cao and M. Yang, J. Phys. B \textbf{36}, 4245 (2003).

\bibitem{zhanglihua}L. H. Zhang, M. Yang, and Z. L. Cao, Phys. A
\textbf{374}, 611 (2007).

\bibitem{wanghf1} H. F. Wang, S. Zhang, and K. H. Yeon, Eur. Phys. J.
D \textbf{56}, 271 (2010).

\bibitem{wanghf2}H. F. Wang, L. L. Sun, S. ZHang, and K.-Y. Yeon,
Quant. Inf. Process. DOI 10.1007/s11128-011-0255-9, (2011)

\bibitem{yildiz} A. Yildiz, Phys. Rev. A \textbf{82}, 012317 (2010).


\bibitem{QND1} K. Nemoto, and W. J. Munro,  Phys. Rev. Lett. \textbf{93}, 250502 (2004).

\bibitem{QND2} S. D. Barrett, P. Kok, K. Nemoto, R. G. Beausoleil, W. J. Munro,
 and T. P. Spiller,  Phys. Rev. A \textbf{71}, 060302 (2005).

\bibitem{shengbellstateanalysis} Y. B. Sheng, F. G. Deng, and G. L.
Long, Phys. Rev. A  \textbf{82}, 032318 (2010).



\bibitem{qubit1} H. Jeong, and N. B. An, Phys. Rev. A \textbf{74},
022104 (2006).

\bibitem{qubit2} G. S. Jin, Y. Lin, B. Wu, Phys. Rev. A
\textbf{75}, 054302 (2007).

\bibitem{qubit3} Y. M. Li, K. S. Zhang, and K. C. Peng, Phys. Rev. A \textbf{77}, 015802
(2008).

\bibitem{he1} B. He, J. A. Bergou, and Y.-H. Ren, Phys. Rev. A \textbf{76}, 032301
(2007); B. He, M. Nadeem, and J. A. Bergou, \emph{ibid}.
\textbf{79}, 035802 (2009); B. He, Y.-H. Ren, and J. A. Bergou,
\emph{ibid.} \textbf{79}, 052323 (2009); J. Phys. B \textbf{43},
025502 (2010).


\bibitem{he2} B. He, Q. Lin, and C. Simon, Phys. Rev. A \textbf{83}, 053826
(2011).
\bibitem{he3} B. He and J. A. Bergou,  Phys.
Rev. A \textbf{78}, 062328 (2008).

\bibitem{lin1} Q. Lin and J. Li, Phys. Rev. A \textbf{79}, 022301 (2009).

\bibitem{lin2} Q. Lin and B. He, Phys. Rev. A \textbf{80}, 042310 (2009); Q. Lin, B. He,
J. A. Bergou, and Y.-H. Ren,\emph{ ibid.} \textbf{80}, 042311
(2009); Q. Lin and B. He, \emph{ibid.} \textbf{80}, 062312 (2009);
Q. Lin and B. He, \emph{ibid.} \textbf{82}, 022331 (2010); Q. Lin
and B. He, \emph{ibid.} \textbf{82}, 064303 (2010).





\bibitem{zhangshou} Q. Guo, J. Bai, L. Y. Cheng, X. Q. Shao, H. F.
Wang, and S. Zhang, Phys. Rev. A \textbf{83}, 054303 (2011).





\bibitem{coffman} V. Coffman, J. Kundu, and W. K. Wootters, Phys. Rev. A
\textbf{61}, 052306 (2000).

\bibitem{plesch}M. Plesch, and V. Bu\u{z}ek, Phys. Rev. A
\textbf{68}, 012313 (2003).

\bibitem{baokui1}B. K. Zhao, F. G. Deng, F. S. Zhang, and H. Y. Zhou,
Phys. Rev. A \textbf{80}, 052106 (2009).

\bibitem{baokui2}B. K. Zhao, and F. G. Deng, Phys. Rev. A \textbf{82},
014301(2010).

\bibitem{concurrrence}W. K. Wootters, Phys. Rev. Lett. \textbf{80}, 2245
(1998).
\bibitem{kok1} P. Kok, W. J. Munro, K. Nemoto, T. C. Ralph, J. P. Dowing, and G.
J. Milburn, Rev. Mod. Phys. \textbf{79}, 135 (2007).

\bibitem{kok2} P. Kok, H. Lee,
and J. P. Dowling, Phys. Rev. A \textbf{66}, 063814 (2002).

\bibitem{hofmann} H. F. Hofmann, K. Kojima, S. Takeuchi,
and K. Sasaki, J. Opt. B \textbf{5}, 218 (2003).


\bibitem{Banacloche}J.Gea-Banacloche, Phys. Rev. A \textbf{81}, 043823 (2010).

\bibitem{Shapiro1} J. H. Shapiro, Phys. Rev. A \textbf{73}, 062305 (2006).

\bibitem{Shapiro2} J. H. Shapiro and M. Razavi, New J. Phys. \textbf{9}, 16 (2007).



\bibitem{weak_meaurement} A. Feizpour, X. Xing, and A. M. Steinberg,
Phys. Rev. Lett. \textbf{107}, 133603 (2011).

\bibitem{oe} C. Zhu and G. Huang, Optics Express, \textbf{19}, 23364 (2011).
\end{thebibliography}
\end{document}